\documentclass[10pt,notitlepage,showpacs,pra,twocolumn]{revtex4-1}
\usepackage{ulem}
\usepackage{amsfonts}
\usepackage{amsmath}
\usepackage{amssymb}
\usepackage{graphicx}
\usepackage{float}
\usepackage{chemarrow}
\usepackage[toc,page,header]{appendix}

\begin{document}

\title{Scalability and high-efficiency of an $(n+1)$-qubit Toffoli gate sphere via blockaded Rydberg atoms}
\author{Dongmin Yu$^{1}$, Yichun Gao$^{1}$, Weiping Zhang$^{2,3}$, Jinming Liu$^{1}$ and Jing Qian$^{1,3,\dagger}$ }
\affiliation{$^{1}$State Key Laboratory of Precision Spectroscopy, Department of Physics, School of Physics and Electronic Science, East China
Normal University, Shanghai 200062, China}
\affiliation{$^{2}$Department of Physics and Astronomy, Shanghai Jiaotong University and Tsung-Dao Lee Institute, Shanghai 200240, China}
\affiliation{$^{3}$Collaborative Innovation Center of Extreme Optics, Shanxi University, Taiyuan, Shanxi 030006, China}

\begin{abstract}
The Toffoli gate serving as a basic building block for reversible quantum computation, has manifested its great potentials in improving the error-tolerant rate in quantum communication. While current route to the creation of Toffoli gate requires implementing sequential single- and two-qubit gates, limited by longer operation time and lower average fidelity. We develop a new theoretical protocol to construct a universal $(n+1)$-qubit Toffoli gate sphere based on the Rydberg blockade mechanism, by constraining the behavior of one central target atom with $n$ surrounding control atoms. Its merit lies in the use of only five $\pi$ pulses independent of the control atom number $n$ which leads to the overall gate time as fast as $\sim$125$n$s and the average fidelity closing to 0.999. The maximal filling number of control atoms can be up to $n=46$, determined by the spherical diameter which is equal to the blockade radius, as well as by the nearest neighbor spacing between two trapped-atom lattices. Taking $n=2,3,4$ as examples we comparably show the gate performance with experimentally accessible parameters, and confirm that the gate errors mainly attribute to the imperfect blockade strength, the spontaneous atomic loss and the imperfect ground-state preparation. In contrast to an one-dimensional-array configuration it is remarkable that the spherical atomic sample preserves a high-fidelity output against the increasing of $n$, shedding light on the study of scalable quantum simulation and entanglement with multiple neutral atoms.

\end{abstract}
\email{jqian1982@gmail.com}
\maketitle
\preprint{}

\section{Introduction}

The long-range nature of the Rydberg-Rydberg interactions in $n$ control atoms can collectively manipulate a single target atom, applying for an $(n+1)$-qubit Toffoli gate \cite{Saffman16}, which has served as a useful protocol for quantum error correction and reversible arithmetic operation in a large number of systems \cite{Cory98,Schindler11,Paetznick13}. When $n=2$ this corresponds to a usual three-qubit Toffoli gate that has found realizations in trapped ions \cite{Monz09,Figgatt19}, superconducting circuits \cite{Fedorov12,Reed12,Baekkegaard19}, superconducting artificial atoms \cite{Zahedinejad15}, single photonic architecture \cite{Lanyon08}, silicon spin qubits \cite{Gullans19} and linear optical systems \cite{Micuda13,Huang17}. For larger $n$ it remains elusive for experimental implementation due to the imperfect manipulation of multiple qubits that requires a composition of single- or two-qubit gates, increasing the complexity of system \cite{Barenco95,Ralph07,Yu13,Biswal19}. 
In contrast to these previous efforts using versatile platforms, a system associated with Rydberg blockade interactions, originally found by M. Lukin {\it et. al.} \cite{Lukin01}, can essentially provide a clean environment to implement an $(n+1)$-qubit Toffoli gate since the single target atom acquires a conditional excitation induced by the competitive excitation of $n$ control atoms within the ``blockade radius''. Here the blockade radius is defined to characterize the maximal value of two-atom separation, within which any one of $n$ control atom's excitation can strongly block the excitation of the unique target atom \cite{Pritchard10,Qian13}.

So far the family of $(n+1)$-qubit Rydberg Toffoli gates, has absorbed attractive attentions for a long time owing to their remarkable properties (excellent quantum coherence and long-range interaction) in large-scale quantum computation and secure communication \cite{Saffman10}. However the implementation of multiqubit Toffoli gates with Rydberg atoms still stays at the theoretical level, lacking of practical developments due to relatively low quantum state initialization and ground-state coherent control \cite{Isenhower10}. Until very recently, the Toffoli gate based on Rydberg blockade was first demonstrated experimentally with atoms trapped in an one-dimensional (1D) array of optical tweezers, leading to an average fidelity $\geq 0.837$ \cite{Levine19}. While such a 1D array configuration may be uneasy for extending into a largely scalable quantum computing network because the next-to-nearest neighbor interaction between long-range atoms is weak, insufficient for blocking the excitation of a middle target atom. The three-body F\"{o}rster resonance among three interacting atoms has been utilized to facilitate interatomic dipole-dipole interactions resulting in a fast and high-fidelity three-qubit Toffoli gate \cite{Beterov18}.

In the current work, we present a new three-dimensional (3D) spherical atomic ensemble for realizing multiqubit Rydberg Toffoli gate based on the blockade, promising for a scalable $(n+1)$-qubit {\it Toffoli sphere} with up to $n=46$ control atoms. Unlike previous proposals utilizing 1D linear or two-dimensional (2D) square arrays to trap atoms \cite{Isenhower11,Gulliksen15,Shi18,Cao18,Graham19}, our model remarkably offers sufficient blockaded energy for arbitrary two atoms inside the sphere, once its diameter is equal to the two-atom blockade radius, leading to a flexible implementation of arbitrary-qubit Toffoli gate. Our scheme is verified to ensure an ultrahigh fidelity output $\sim0.999$ accompanied by a very short gate time $\sim 125n$s, with the help of only five common $\pi$ pulses that is fully independent of the number of control qubits $n$ \cite{Rasmussen19}.

Taking $n=2$ as an instance, the major contributions to the gate error come from the imperfect blockade and finite Rydberg state lifetime. The former has manifested as a routine error considered for all blockaded gates \cite{Saffman05,Zhang12,Shi17}. Here a more precise scaling relation $\mathcal{E}_b\sim0.5(\Omega_0/V_0)^{1.97}$ is numerically supported. The latter will cause an inevitable spontaneous decay error identifying an exact linear scaling as $\mathcal{E}_{\Gamma}\sim\Gamma/\Omega_0$ for the pulse duration $\propto1/\Omega_0$ \cite{Petrosyan17}. The imperfect ground state initialization to the gate fidelity is complementarily explored according to the experimental parameters, providing us a straightforward route to realize higher fidelities of multiqubit Rydberg Toffoli gate with a stronger blockade strength, longer lifetimes, and more stable initial preparations. In contrast to a traditional 1D array protocol, our system benefits from a preserved strong blockade covering any atom inside the Toffoli sphere, robustly ensuing a scalable multiqubit quantum computation.

\section{Basic three-atom Toffoli Gate}

\subsection{Protocol of a $(2+1)$-qubit gate}

 \begin{figure}
\includegraphics[width=3.6in,height=2.4in]{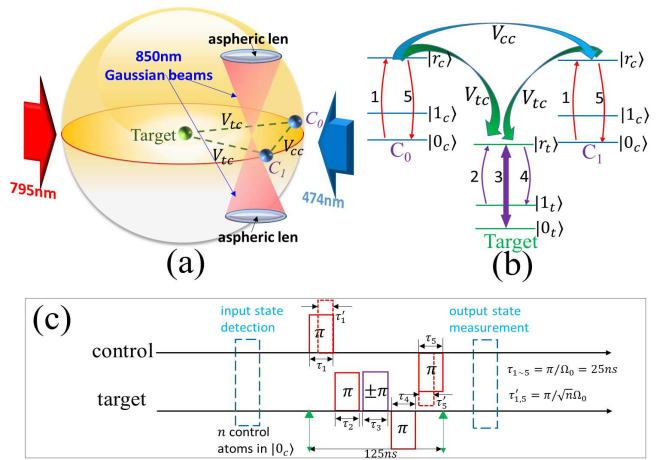}
\caption{(color online). (a) Schematic diagram of simplest $(2+1)$-qubit Rydberg Toffoli gate configuration that can be extended to a 3D spherical structure [also see Fig.\ref{sphere}a]. The only target atom $T$ (green circle) placed in the spherical center is surrounded by 2 identical control atoms $C_0$ and $C_1$ (blue circles), yielding a $(2+1)$-qubit gate. Here $V_{tc}$ and $V_{cc}$ describe the nearest target-control and control-control Rydberg-Rydberg interactions, respectively. Ground rubidium atoms arranged in a triangular lattice are globally driven to Rydberg states by using a two-photon transition via counter-propagating 795nm and 474nm laser beams, coupling the transition of $5S_{1/2}\to5P_{1/2}\to82D_{3/2}$. The intermediate state $5P_{1/2}$ is largely detuned. This triangular microscopic optical trap (triangular lattice) is formed by focusing 850nm Gaussian lasers with a high-numerical-aperture aspheric lens \cite{Nogrette14}. (b) Detailed interpretation of three atomic energy levels, interacting with five $\pi$-pulses $\Omega_{1\sim5}$ denoted as $1\sim5$. (c) Real timing diagram of $\Omega_{1\sim5}$ with equal amplitude $\Omega_0$ and duration $\tau_{1\sim5}=\pi/\Omega_0$. When more than two input control atoms (denoted by $n$) are prepared in state $|0_c\rangle$ by detection, the duration of $\Omega_{1,5}$ becomes $ \tau_{1,5}^{\prime}= \pi/(\sqrt{n}\Omega_0)$ with the effective pulse amplitude $\sqrt{n}\Omega_0$ for preserving the $\pi$-pulse area. The total gate pulse time persists a constant value 125$n$s. Noting $\Omega_4 =-\Omega_2 $, $\Omega_5=-\Omega_1$ enabled by a $\pi$-phase difference for the two respective transitions. }
\label{mod}
\end{figure}

The basic idea of our Toffoli sphere is enabled by a $(2+1)$-qubit gate, as illustrated in Fig. \ref{mod}(a), consisting of three atoms which involve two control atoms $C_0$, $C_1$ and one target atom $T$, trapping in {\it e.g.} a tunable triangular optical lattice \cite{Jo12}. Arbitrary geometries of 2D arrays can be obtained by imprinting a phase map with the spatial light modulator \cite{Barredo14}. It is assumed that, each atom contains two hyperfine ground states $|0_{c_0,c_1,t}\rangle$, $|1_{c_0,c_1,t}\rangle$ and one Rydberg state $|r_{c_0,c_1,t}\rangle$, experiencing the optical excitation and de-excitation transitions by five $\pi$-pulses $\Omega_{1\sim 5}$. When any two of three atoms are coupled to Rydberg levels simultaneously it leads to a Rydberg energy shift defined by $V_{cc}$ for two control atoms and by $V_{tc}$ for one control and one target atoms. Without loss of generality all control atoms are considered to be identical by safely ignoring the subscripts {\it 0,1,2, ...} of $c$ throughout the paper.

\begin{widetext}

\begin{figure}
\centering
\includegraphics[width=5.0in,height=2.8in]{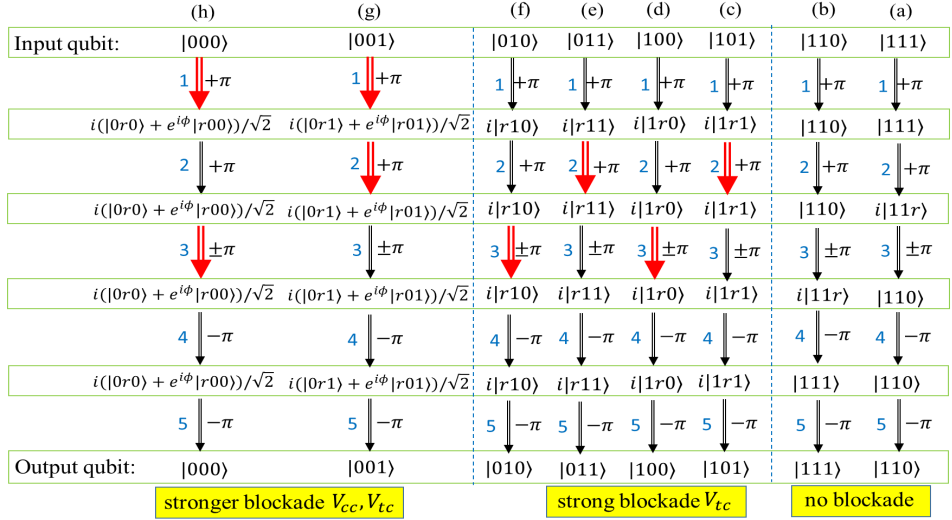}
\caption{(color online). Complete procedure of a $(2+1)$-qubit Toffoli gate, for all input and output qubits $|\alpha\alpha\beta\rangle$. 
{\it Case 1:} (a-b) No blockaded energy is required when two control atoms are both in the idle state $|11\rangle$. {\it Case 2:} (c-f) While one of two control atoms is prepared in $|1\rangle$, it can fully prohibit the target atom's excitation owing to the blockade energy $V_{tc}$ between control and target atom pairs. {\it Case 3:} (g-h) When both control qubits are initially in $|00\rangle$ it remarkably induces a superposition state by collectively exciting one control atom to the Rydberg state $|r_c\rangle$ due to the blockaded energy $V_{cc}$, breaking the excitation of the third target atom if $V_{tc}$ is sufficient at the same time. Here the steps involving the Rydberg blockade mechanism are denoted by red arrows. The area of all square pulses is $\pi$ and the sign ``$\pm$'' stand for different excitation or de-excitation directions and $k=1\sim5$ denote the pulse $\Omega_k$.}
\label{procedure}
\end{figure}

\end{widetext}

Provided $\pi$ pulse sequences $\Omega_{1\sim 5}$ like Fig. \ref{mod}(c) the corresponding atom-field interactions are demonstrated in (b) where $\Omega_{1,5}$ makes the transition of $|0_c\rangle\rightleftarrows|r_c\rangle$ for control atoms, $\Omega_{2,4}$ makes the $|1_t\rangle\rightleftarrows|r_t\rangle$ transition for the target atom, and $\Omega_{3}$ results in a reversible target population exchange between $|0_t\rangle$ and $|r_t\rangle$. Remember $|1_c\rangle$ is an idler for control atoms. The detailed procedure for gate performance depending on the Rydberg blockade mechanism can be understood by following the diagram of Fig. \ref{procedure} where we write the basis state as $|\alpha\alpha\beta\rangle$ with $\alpha\in(0_c,1_c)$, $\beta\in(0_t,1_t)$. Noting here two control atoms are having same interactions with the target atom.

Depending on different initialization of two control atoms, whether $|\alpha\rangle=|0_c\rangle$ or not, the whole gate procedure can be classified into three {\it Cases}:

{\it Case 1:} When two control qubits are initialized in $|11\rangle$, see Fig.\ref{procedure}(a-b), no blockade effect appears since $|1_c\rangle$ is idle. The target atom initialized in $|0_t\rangle$ or $|1_t\rangle$ will solely experience a state swapping along 
\begin{eqnarray}
|0_t\rangle\xrightarrow{\Omega_3}i|r_t\rangle\xrightarrow{-\Omega_4}|1_t\rangle \nonumber\\
|1_t\rangle\xrightarrow{\Omega_2}i|r_t\rangle\xrightarrow{-\Omega_3}|0_t\rangle \nonumber
\end{eqnarray}
due to the apply of $\pi$-pulses. The negative sign means a $\pi$ phase difference for Rabi frequencies in the respective transitions.

{\it Case 2:} When only one of the two control atoms is initialized in $|0_c\rangle$, {\it i.e. $|01\beta\rangle$ or $|10\beta\rangle$}, it can be excited to state $|r_c\rangle$ by pulse 1, arising a strong blockade energy $V_{tc}$ to prohibit the excitation of the target atom. As a consequence the target atom sustains on the initial state accompanied by the unique control atom, experiencing a state transition of
\begin{equation}
|0_c\rangle\xrightarrow{\Omega_1}i|r_c\rangle\xrightarrow{-\Omega_5}|0_c\rangle \nonumber
\end{equation}
which gives rise to same output qubit state as the initialization, as shown in Fig.\ref{procedure}(c-f).

{\it Case 3:} While two control atoms are prepared in state $|0_c0_c\rangle$ it enables a simultaneous excitation to double Rydberg state $|r_cr_c\rangle$ via pulse 1. However in the presence of a strong interaction $V_{cc}$ arising a big amount of shifted energy to $|r_cr_c\rangle$ one may expect a collective entangled state $|\Psi_{+}\rangle$ like
\begin{equation}
|\Psi_{+}\rangle=(1/\sqrt{2})(|0_cr_c\rangle+e^{i\phi}|r_c0_c\rangle)
\end{equation}
with only one control qubit being coherently excited \cite{Gaetan09}. To this end the effective induced Rabi frequency between $|0_c0_c\rangle$ and $|\Psi_{+}\rangle$ becomes $\sqrt{2}\Omega_0$ leading to the requirement of a shorter pulse duration $\tau_{1,5}^{\prime}=\pi/\sqrt{2}\Omega_0$ for preserving the $\pi$-pulse property.
Here the key ingredient lies in, once any one control qubit is collectively excited due to a big blockaded energy $V_{cc}$, it is able to prevent the excitation of the target atom no matter it is initialized in $|0_t\rangle$ or $|1_t\rangle$, as long as the interatomic interaction $V_{tc}$ is sufficient ($r_{tc}\leq R_{b}$, here $r_{tc}$ is the target-control distance, $R_{b}$ is the two-atom blockade radius). Therefore with the initial qubit state $|00\beta\rangle$ [see Fig.\ref{procedure}(g-h)], the target atom can maintain its initialization, and at the same time two control atoms undergo collective excitation and de-excitation, returning back to the initial state $|00\rangle$ after pulse 5. 

By then we principally accomplish a realization of the three-qubit Toffoli blockaded gate in a 2D triangular lattice, allowing all input-output transformations obeying the map of :
\begin{eqnarray}
&|\Phi_{i}\rangle&=\{|000\rangle,|001\rangle,|010\rangle,|011\rangle,|100\rangle,|101\rangle,|110\rangle,|111\rangle\} \nonumber\\
&\xrightarrow{\Omega_{1\sim5}}& \nonumber\\
&|\Phi_{f}\rangle&=\{|000\rangle,|001\rangle,|010\rangle,|011\rangle,|100\rangle,|101\rangle,|111\rangle,|110\rangle\} \nonumber
\end{eqnarray}
in which the importance of blockade strengths $V_{cc}$ and $V_{tc}$ has been pointed out.

\subsection{High-quality gate performance}

We proceed to exploit the gate performance by numerically solving the time evolution of master equation via fourth-order Runge-Kutta (RK) algorithm: $\hat{\rho}(t)=-i[\hat{\mathcal{H}}_0(t)+\hat{\mathcal{H}}_I,\hat{\rho}]+\hat{\mathcal{L}}_{c0}+\hat{\mathcal{L}}_{c1}+\hat{\mathcal{L}}_{t}$ with $\hat{\mathcal{H}}_0$ the time-dependent atom-light interactions
\begin{eqnarray}
\hat{\mathcal{H}}_0(t)&=&\frac{1}{2}\{\Omega_1\hat{\sigma}_{0_cr_c}+\Omega_5\hat{\sigma}_{r_c0_c}+\Omega_2\hat{\sigma}_{1_tr_t}+\Omega_4\hat{\sigma}_{r_t1_t}  \nonumber \\
&+&\Omega_3(\hat{\sigma}_{0_tr_t}+\hat{\sigma}_{r_t0_t})\}
\end{eqnarray}
and $\hat{\mathcal{H}}_I$ the atom-atom vdWs interactions
\begin{equation}
\hat{\mathcal{H}}_I=2V_{tc}|r_cr_t\rangle\langle r_cr_t|+V_{cc}|r_cr_c\rangle\langle r_cr_c|
\end{equation}
Here the transition operator is defined by $\hat{\sigma}_{mn}=|m\rangle\langle n|$ and state $|1_c\rangle$ is kept to be an idler. $\hat{\mathcal{L}}_{c,t}$ standing for the spontaneous emission from the Rydberg level $|r_{c,t}\rangle$, takes the form of
\begin{equation}
\hat{\mathcal{L}}_k[\hat{\rho}(t)]=\frac{1}{2}\sum_j^{2}[2\hat{\mathcal{L}}_{j,k}\hat{\rho}\hat{\mathcal{L}}_{j,k}^{\dagger}-(\hat{\mathcal{L}}_{j,k}^{\dagger}\hat{\mathcal{L}}_{j,k}\hat{\rho}+\hat{\rho}\hat{\mathcal{L}}_{j,k}^{\dagger}\hat{\mathcal{L}}_{j,k})]
\end{equation}
with $k\in(c,t)$ and $\hat{\mathcal{L}}_{1,k}=\Gamma|1_k\rangle\langle r_k|$, $\hat{\mathcal{L}}_{2,k}=\Gamma|0_k\rangle\langle r_k|$. $\Gamma$ denotes the spontaneous decay rate of $|r_{c,t}\rangle$. Considering the gate of three atoms that the entire density operator $\hat{\rho}$ becomes a $3^3\times 3^3$ matrix, and the decay rate $\Gamma$ will cause the leakage of population onto other undesired states lowering the fidelity as long as state $|r_{c,t}\rangle$ is involved. As far as we know, the three-qubit Toffoli gate in neutral atoms mediated by Rydberg excitation has recently reported \cite{Levine19} although the theoretical pursuit of neutral-atom Toffoli gate has experienced a long history. Here our target is achieving a reliable and simpler way for an arbitrary multiqubit Toffoli gate via Rydberg blockade, simultaneously benefiting from a fast operation time $\sim125n$s and high-fidelity $\sim$0.999. 

Consider three identical $^{87}$Rb atoms, individually captured with microscopic optical tweezers in a 2D triangular lattice, with the hyperfine ground states $|0(1)\rangle = |5s_{1/2},f = 1(2)\rangle$ and Rydberg state $|r\rangle = 82D_{3/2},m_J = 3/2\rangle$. The lattice spacing has a wide change from 0.96$\mu$m to $11.2\mu$m with real-time control of the periodicity of lattice \cite{Li08}, providing us opportunities to flexibly adjust the two-atom separations $r_{tc},r_{cc}$ in the design. In addition the anisotropy of the van der Waals (vdWs) interactions of $|nD\rangle$ Rydberg states causes a tunable two-atom Rydberg interaction energy that depends on controlling the angle between internuclear axis and quantization axis, as demonstrated by experiment \cite{Barredo14} which may be an alternative way for achieving tunable interactions in a certain range.
For $|r\rangle = |82D_{3/2}\rangle$ the experimental coefficient $C_6$ qualifying the vdWs interaction between two isolated-trapped Rydberg atoms is $C_6/2\pi = 1400$GHz$\mu m^6$, where the blockade radius was measured around $R_b\sim$4.0$\mu$m \cite{Beguin13}. The state $|r\rangle$ has a decay rate $\Gamma/2\pi = 1.5$kHz at zero temperature \cite{Beterov09}. We perform a reasonable effective Rabi frequency $\Omega_0/2\pi = 20$MHz exciting atoms from qubit states $|0_c\rangle$, $|0(1)_t\rangle$ to the Rydberg states $|r_{c}\rangle$ and $|r_{t}\rangle$ via a two-photon process, giving to the
minimal time for gate operation $T=5\pi/\Omega_0 = 125ns$, and the ratio of decay is $\Gamma/\Omega_0=7.5\times10^{-5}$.
Note that if the input state is $|00\beta\rangle$ by detection, one should self-consistently decrease the pulse durations $\tau_{1,5}^{\prime}=\pi/\sqrt{2}\Omega_0<\tau_{1,5}$ in {\it Case 3} however the whole gate time $T$ is unvaried. For preserving the strong blockade between nearest neighbors, we use $V_0 = V_{tc(cc)} =2\pi\times240$MHz$\gg\Omega_0$ which is verified to be a threshold of Rydberg blockade [see Sec. \rm{III}B], leading to a more precise value of blockade radius that is $R_{b}=(C_6/V_0)^{1/6}=4.24\mu$m.

In our calculation the time interval in RK algorithm is $\Delta\tau\sim 0.1n$s and the measured variables are
\begin{eqnarray}
\mathcal{F}_{\alpha\alpha\beta} =  _{\alpha\alpha\beta}\langle \Phi_{f}| \rho(t=t_{det})|\Phi_f\rangle _{\alpha\alpha\beta}  \nonumber\\
\mathcal{F}_0 =\frac{1}{\mathcal{N}} \text{Tr}[\sqrt{\rho_{et}}\rho(t=t_{det})\sqrt{\rho_{et}}]^{1/2}
\end{eqnarray}
where $\mathcal{F}_{\alpha\alpha\beta} $ is an element fidelity (EF) of each qubit $|\alpha\alpha\beta\rangle$, $\mathcal{F}_0$ describes the average fidelity (AF) of the gate quality by comparing the real final state $\rho(t)$ at $t=t_{det}$ after time evolution with the etalon state $\rho_{et}$ of an ideal Toffoli gate, averaging over all input states \cite{Nielsen11}. The measurement time is $t_{det}=125n$s, $\mathcal{N}=8$ denotes the number of input states.

\begin{figure}
\centering
\includegraphics[width=3.4in,height=2.8in]{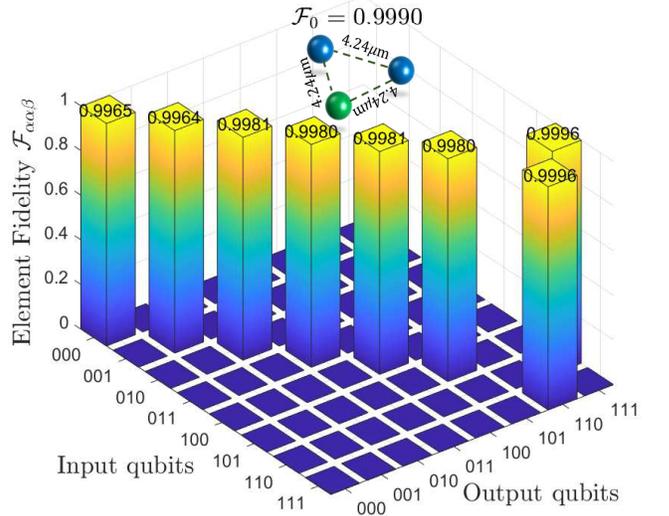}
\caption{(color online). In a $(2+1)$-qubit Toffoli gate scheme as modeled in the inset, the measured truth table for the EF $\mathcal{F}_{\alpha\alpha\beta}$ of arbitrary input and output qubit states $|\alpha\alpha\beta\rangle$. The AF attains as high as $\mathcal{F}_0=0.9990$. Relevant parameters are described in the text.}
\label{aver}
\end{figure}

At the satisfactory of sufficient blockade strengths we expect a high-fidelity transfer of input states to the desired output states as shown in Figure \ref{aver} which exhibits the full gate performance by calculating arbitrary EFs, as well as the AF of system, denoted by $\mathcal{F}_{\alpha\alpha\beta}$ and $\mathcal{F}_0$. Perfectly agreeing with the theoretical predictions of the three {\it Cases}, our result demonstrates that the inputs $|110\rangle$ and $|111\rangle$ benefit from a maximal fidelity value $\mathcal{F}_{11\beta}\sim0.9996$ due to its complete independence on the blockade effect that is only affected by the decay error $\sim (7\pi/4)(\Gamma/\Omega_0)=4\times10^{-4}$ \cite{Zhang12}. While if one of two control atoms is initialized in $|0_c\rangle$ the scheme suffers from a target-control blockade $V_{tc}$, leading to a relatively lower value $\mathcal{F}_{01(10)\beta}\sim0.9980(1)$. As expected the lowest value $\mathcal{F}_{00\beta}\sim0.9964(5)$ appears in the presence of input qubits $|000\rangle$ and $|001\rangle$ since the control-control and target-control blockades both play crucial roles there. The reason behind can be roughly understood by loading the concept of blockade error from a residual double excitation probability due to imperfect blockade, typically scaling as $\propto \Omega_0^2/2V_0^2\sim10^{-3}$ by theoretical analysis \cite{Zhang12,Petrosyan17,Shi17} and experimental confirmation \cite{Isenhower10}, arising a reduction of the gate fidelity. Luckily, it is shown that the average value $\mathcal{F}_0$ can reach as high as 0.9990 to our $(2+1)$-qubit Toffoli gate within a very short gate operation time $125n$s, where all possible errors coming from imperfect blockade strength, inevitable spontaneous losses of Rydberg levels and imperfect ground state initialization will be left for discussions in the next section.

\section{Gate error discussion}

\subsection{Blockade threshold and blockade error}

To our knowledge the intrinsic blockade error limited by finite Rydberg blockade shift inevitably exists in blockaded gates, which is scaled by $\sim0.5(\Omega_0/V_0)^2$ as analyzed by Saffman and Walker originally \cite{Saffman05}. Owing to the apply of different control-control and target-control interactions, the realistic blockade error may suffer from the imperfect strengths of $V_{tc}$ and $V_{cc}$ at the same time, bringing modifications to this scaling relation.

In the following we carry out an exploration of the blockade strength threshold, paying special attention to the relative ratio between $V_{tc}$ and $V_{cc}$, for obtaining the minimal values of blockade strengths. Similarly we define a partial average fidelity (PAF) covering part of input states, as
\begin{eqnarray}
\mathcal{F}_{0,01} =\frac{1}{4} \text{Tr}\sqrt{\sqrt{\rho_{et,01}}\rho_{01}(t=t_{det})\sqrt{\rho_{et,01}}} \\
\mathcal{F}_{0,00} =\frac{1}{2} \text{Tr}\sqrt{\sqrt{\rho_{et,00}}\rho_{00}(t=t_{det})\sqrt{\rho_{et,00}}}
\end{eqnarray}
with $\mathcal{F}_{0,01}$ for four states $|01\beta\rangle$ and $|10\beta\rangle$ in {\it Case 2} that are only affected by the target-control blockade $V_{tc}$, and with $\mathcal{F}_{0,00}$ for two states $|00\beta\rangle$ in {\it Case 3} affected by stronger blockade $V_{tc}$ and $V_{cc}$. Accordingly, $\rho_{et,01}$ and $\rho_{et,00}$ describe the etalon states of $\{|010\rangle,|011\rangle,|100\rangle,|101\rangle\}$ and $\{|000\rangle,|001\rangle\}$, $\rho_{01}(t)$ and $\rho_{00}(t)$ are the corresponding real-time density matrix element of them. Here we ignore $|11\beta\rangle$ because no blockade exists if the input state is $|11\beta\rangle$, arising the PAF for $|11\beta\rangle$ limited by a decay error only. However, we believe that $\mathcal{F}_{0,01}$ and $\mathcal{F}_{0,00}$ will be sensitively modified by using adjustable target-control and control-control interactions, leading to an interaction-dependent oscillating behavior for the final output state probability.

\begin{figure}
\centering
\includegraphics[width=3.5in,height=1.4in]{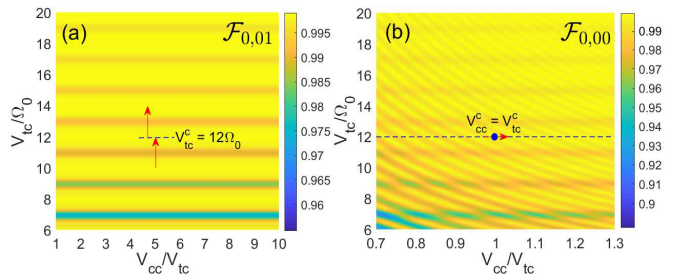}
\caption{(color online). (a-b) The partial average fidelities $\mathcal{F}_{0,01}$ and $\mathcal{F}_{0,00}$ versus tunable interactions denoted by the ratios $V_{cc}/V_{tc}$ and $V_{tc}/\Omega_0$, with respect to {\it Case 2} and {\it Case 3}. 
The black-dashed line or red arrow calibrates the threshold value of blockade by numerical ways where critical values $V_{tc}^c/\Omega_0 = 12$ and $V_{cc}^{c}/V_{tc}^c = 1.0$ are found. Here $\Omega_0 = 2\pi\times20$MHz.}
\label{con}
\end{figure}

Figure \ref{con} globally shows the oscillating behavior of $\mathcal{F}_{0,01}$ and $\mathcal{F}_{0,00}$ where a continuous change for the ratios of $V_{cc}/V_{tc}$ and $V_{tc}/\Omega_0$, instead of their absolute values, is used for presenting the least requirement of blockade strengths. It is obvious that $\mathcal{F}_{0,01}$ [(a)] only depends on $V_{tc}/\Omega_0$, perfectly agreeing with our interpretations in {\it Case 2} that only the target-control interaction plays roles. As increasing $V_{tc}/\Omega_0$ (keeping $\Omega_0$ a constant) $\mathcal{F}_{0,01}$ reveals an amplitude-decaying oscillation tending towards a saturation as $V_{tc}$ is sufficiently large. The threshold value for $V_{tc}/\Omega_0$ can be numerically estimated in Fig. \ref{con}(a) by comparing two maximal values in two nearby oscillating periods, allowing the difference smaller than $<10^{-4}$. According to the criterion we solve the threshold value $V_{tc}^{c}\approx 12\Omega_0  =2\pi\times 240$MHz, arising a critical blockade radius for the target-control blockade, which is $R_{b} = 4.24\mu$m.

In addition, seen from {\it Case 3} the PAF $\mathcal{F}_{0,00}$ affected by two blockade mechanisms, will expectedly reveal intenser interaction-dependent oscillations when $V_{cc}/V_{tc}$ or $V_{tc}/\Omega_0$ varies, attributing to the constraints of stronger blockade condition. Based on the analysis of target-control blockade threshold $V_{tc}^{c}$ we only adjust $V_{cc}/V_{tc}$, denoted by the black-dashed line in (b), in order to see the interaction-dependent behavior of $\mathcal{F}_{0,00}$ for obtaining a minimal value of $V_{cc}^{c}$. A similar saturation effect with a decaying-oscillation amplitude is found in $\mathcal{F}_{0,00}$ as $V_{cc}/V_{tc}$ increases, verifying the importance of determining a blockade threshold $V_{cc}^{c}$. A same numerical way is implemented by comparing the difference of maximum in nearby oscillating periods, letting it below $10^{-4}$, arising $V_{cc}^c=V_{tc}^c$. The resulting blockade radius for the control-control and target-control atoms is also same.

 Finally we obtain a set of experimental parameters for achieving this high-quality three-qubit Toffoli gate in a real  implementation, which are 
\begin{eqnarray}
&\Omega_0/2\pi&=20\text{MHz},V_0=V_{cc}^{c}/2\pi=V_{tc}^{c}/2\pi =240\text{MHz} \nonumber\\
&R_{b}&=4.24\mu m, \Gamma/2\pi=1.5\text{kHz}, T=125ns. \label{pa}
\end{eqnarray}

\begin{figure}
\centering
\includegraphics[width=3.4in,height=2.6in]{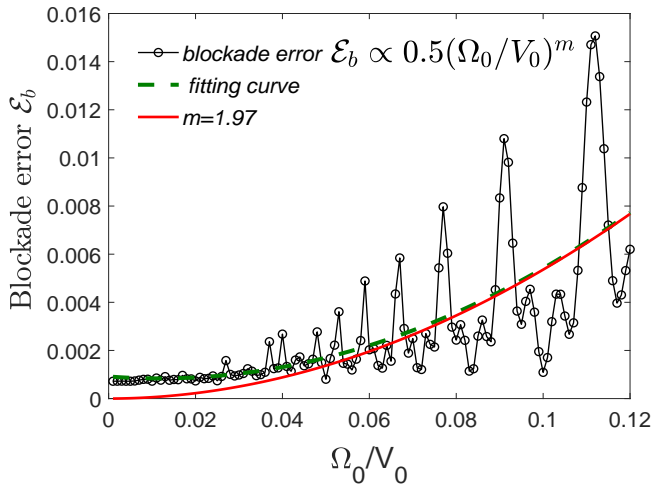}
\caption{(color online). Blockade error $\mathcal{E}_b$ (black-solid with circles) versus the varying of $\Omega_0/V_0$, exactly fitted numerically (green-dashed) under the least square criterion. A theoretical function (red-solid) is used to rescale the numerical data by $\mathcal{E}_b\propto0.5(\Omega_0/V_0)^{1.97}$. Here $V_0 = V_{cc}=V_{tc}$. }
\label{berror}
\end{figure}

Based on which we are able to re-modify the scaling relation of blockade error $\mathcal{E}_b$ defined by $\mathcal{E}_b = 1-\mathcal{F}_{0}$ with respect to the ratio $\Omega_0/V_0$, as shown in  Figure \ref{berror}. Generally speaking, the blockade error $\mathcal{E}_b$ intrinsically comes from imperfect blockade strengths [red arrows in Fig.\ref{procedure}] which becomes worse as $V_0$ decreases. Therefore $\mathcal{E}_b$(black-solid with circles) is observed to continuously enhance accompanied by strong oscillations as increasing $\Omega_0/V_0$, numerically fitted by a green-dashed curve via least-squares criterion. Previous work has proposed a power function $\mathcal{E}_b\sim0.5(\Omega_0/V_0)^2\approx10^{-3}$ to rescale this enhancement \cite{Saffman05}. Here we suggest a more universal function form $(\Omega_0/V_0)^m$ for fitting this curve in order to obtain a more precise value $m$. It implies that $m=1.97$ (red-solid) is a reasonable number to rescale the average data in our scheme, by which the blockade error with parameters in Eq.(\ref{pa}) can be estimated to be $\mathcal{E}_b \sim0.5(\Omega_0/V_0)^{1.97} = 3.7\times10^{-3} $.

\subsection{Error from spontaneous decay}

Furthermore, we introduce $\mathcal{E}_{\Gamma} =1-\mathcal{F}_0$ standing for the decay error and exploit its relation to the
spontaneous decay rate $\Gamma$ of Rydberg states. Ideally, the finite lifetime of highly-excited Rydberg states limited by the decay rate must scale as $1/\Gamma\propto n^3$, accompanying by the fact that a smaller Rabi frequency $\Omega_0$ leads to a longer duration $\tau\sim\pi/\Omega_0$ on Rydberg levels, giving rise to $\mathcal{E}_{\Gamma}\propto\alpha(\Gamma/\Omega_0)$ [$\alpha$ is a coefficient]. Next we will study this linear scaling relation by numerical tools.

\begin{figure}
\centering
\includegraphics[width=3.4in,height=1.4in]{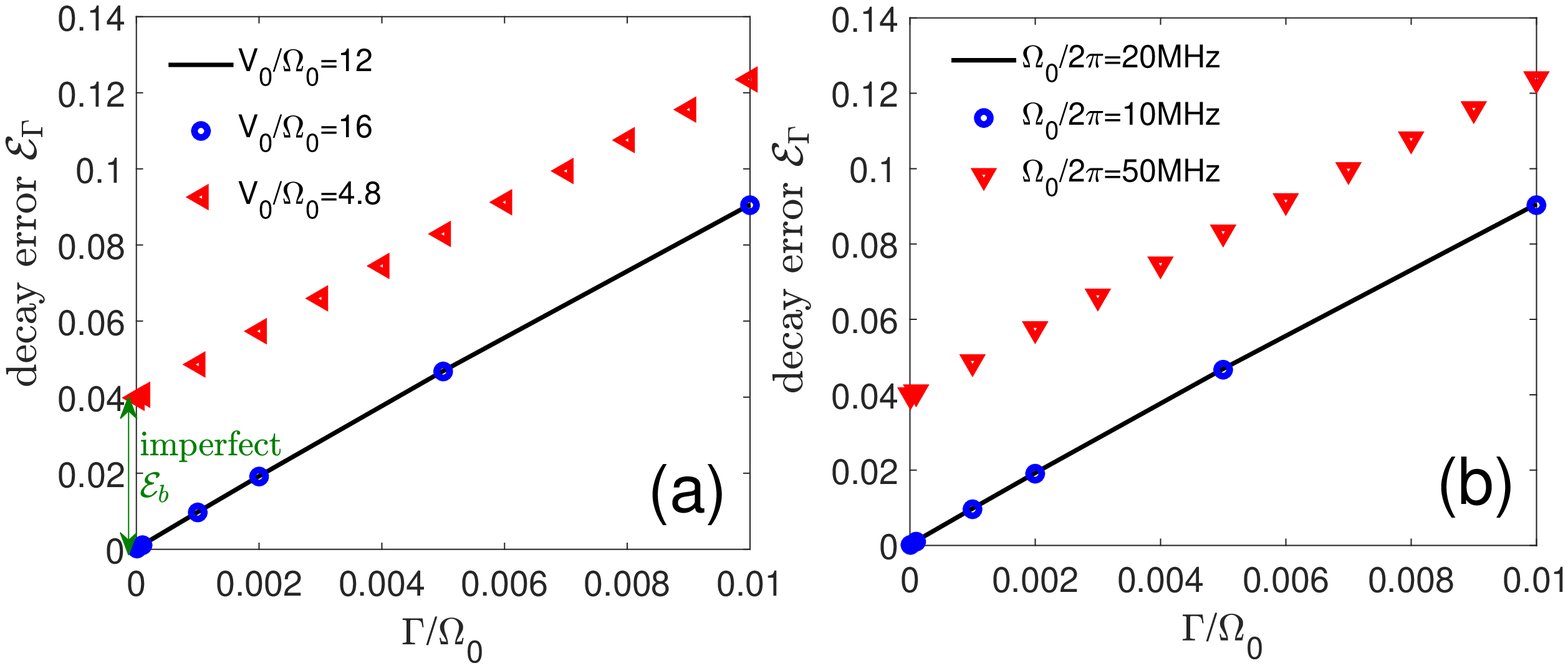}
\caption{(color online). Decay error $\mathcal{E}_{\Gamma}$ stemming from the finite lifetime of Rydberg levels, versus the variation of spontaneous decay rate $\Gamma$. (a) $\mathcal{E}_{\Gamma}$ under different vdWs interaction strengths $V_0/\Omega_0=12$(black-solid, threshold), $4.8$(red triangles), $16$(blue diamonds) where $\Omega_0=2\pi\times20$MHz; (b) $\mathcal{E}_{\Gamma}$ under different absolute optical Rabi frequencies $\Omega_0/2\pi=20$MHz(black-solid), $10$MHz(blue diamonds) and $40$MHz(red triangles) where the blockaded energy is $V_0=2\pi\times 240$MHz. }
\label{decay}
\end{figure}

Figure \ref{decay}(a) plots $\mathcal{E}_{\Gamma}$ versus $\Gamma/\Omega_0$ by tunably adjusting the vdWs interaction strengths $V_0=(12,16,4.8)\Omega_0$ where $\Omega_0/2\pi=20$MHz is fixed. It occurs an exact linear increase for the decay error $\mathcal{E}_{\Gamma}$ as $\Gamma/\Omega_0$ increases from the starting point $\mathcal{E}_{\Gamma}\to 0$, $\Gamma\to 0$, as long as $V_0$ is larger than its threshold value $12\Omega_0$. However an imperfect blockade under $V_0/\Omega_0<12$ (here $V_0/\Omega_0=4.8$, red triangles) will arise an upward shift of this linear curve meaning a poor decay error, yet its slope preserves a constant. A special point where $\Gamma\to0$ for $V_0/\Omega_0=4.8$ stands for the imperfect blockade error around $0.04$, being one order of magnitude larger than the perfect blockade error $\mathcal{E}_b \sim0.0037$.

Also in Fig. \ref{decay}(b), converting Fig. \ref{decay}(a) into another frame by displaying the relation of $\mathcal{E}_{\Gamma}$ and $\Gamma/\Omega_0$ for different absolute Rabi frequencies $\Omega_0$ may give similar results. First an agreeable linear scaling is found, stressing the importance of strong blockade condition $V_0/\Omega_0\geq12$ ($\Omega_0/2\pi\leq20$MHz) in the reduction of decay error $\mathcal{E}_{\Gamma}$. However when $\Omega_0/2\pi=50$MHz ($V_0/\Omega_0=4.8$) it appears a same upward shift for the scaling line as in (a), meaning that even $\Gamma\to0$ there exists an inevitable blockade error from the imperfection of blockade strength. The imperfect blockade error can also be absorbed from the shifted energy on axis $y$ when $\Gamma=0$ since the decay error reduces to zero there. In our calculations the real decay error is estimated to be $\mathcal{E}_{\Gamma}=\alpha(\Gamma/\Omega_0)^1\sim 4.0\times10^{-4}$ [$\alpha = 7\pi/4$ is obtained from Ref. \cite{Zhang12}], perfectly confirming the results of EFs: $\mathcal{F}_{110}=\mathcal{F}_{111}=0.9996$ which are only affected by the intrinsic decay error.

\subsection{Error to imperfect ground-state preparation}

Errors due to the effects of blockade imperfection and spontaneous decays are intrinsic to the blockade gate operation. Besides there exists another artificial error that mainly depends on the experimental technology of initial state preparation. In fact it is impossible to implement an exact accurate preparation of quantum states $|\alpha\alpha\beta\rangle$ with unit efficiency, however the ground state fidelity has been deeply improved by current trapping and detecting tools in experiment.

Taking the input state $|000\rangle$ as an example, we assume the control atom $C_0$ has a $\epsilon_1$-probability in state $|0_c\rangle$ and $(1-\epsilon_1)$-probability in state $|1_c\rangle$. Similarly this tunable coefficient for the control atom $C_1$ and target atom $T$ is respectively $\epsilon_2$ and $\epsilon_3$. Noting that $\epsilon_i$ is a random number generated from a random number generator, ensuing the product $\epsilon_1\epsilon_2\epsilon_3\in[0.955,0.990]$ from experimental facts \cite{Levine18}. To this end if the original input qubit is purely $|000\rangle$ the actual input state becomes a superposition state consisting of full eight bare states $|\alpha\alpha\beta\rangle$. Nevertheless except for $|000\rangle$ whose existing probability is $\epsilon_1\epsilon_2\epsilon_3$ is close to 1.0, the probabilities of other seven states are all very poor owing to its probability proportional to a tiny value $(1-\epsilon_i)$.

\begin{figure}
\centering
\includegraphics[width=3.4in,height=2.6in]{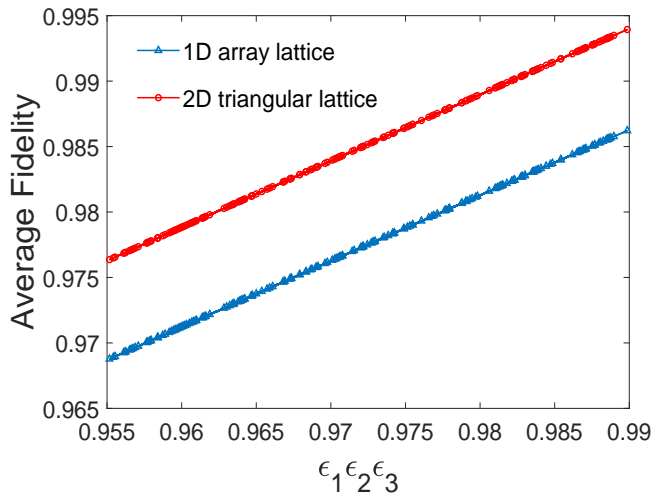}
\caption{(color online). A comparison for the average fidelity $\mathcal{F}_0$ of three-qubit Toffoli gate in 1D array (blue line with circles) and 2D triangular lattices (red line with triangles), versus the change of imperfect probabilities for initial-state preparation, denoted by $\epsilon_1\epsilon_2\epsilon_3$ with $\epsilon_i$ the probability of each initial qubit $\{|\alpha\rangle$, $|\alpha\rangle$, $|\beta\rangle\}$ respectively. Based on experimental results we choose the product $\epsilon_1\epsilon_2\epsilon_3\in[0.955,0.990]$ where $\epsilon_i$ is created randomly.}
\label{initi}
\end{figure}

In Figure \ref{initi} we set $\epsilon_1\epsilon_2\epsilon_3$ as a tunable variable that changes randomly between 0.955 and 0.990, representing the imperfection of initialization, where every $\epsilon_i$ is generated randomly from a similar range. It is clear that the AF $\mathcal{F}_0$ linearly increases with the product of $\epsilon_1\epsilon_2\epsilon_3$, from $\mathcal{F}_0=0.9764$ at $\epsilon_1\epsilon_2\epsilon_3=0.955$ to $\mathcal{F}_0=0.9939$ at $\epsilon_1\epsilon_2\epsilon_3=0.990$. Also this linear scaling can be identified in the model of a 1D three-atom array in which the AF also exhibits a linear increase as increasing $\epsilon_1\epsilon_2\epsilon_3$. Here our proposed scheme (2D triangular lattice) is superior to the former 1D-array type Toffoli gate by an enhanced fidelity of $\sim0.0076$, mainly caused by the stronger blockade strength preserved among three identical atoms in a triangular lattice. While in a 1D array lattice the next to nearest (NN) control-control interaction is longer range and weak, insufficient for blocking the complete excitation of the target atom, arising a poor gate fidelity yield. It is predicted that this initialization error will be accumulated as the number of gate atoms increases, bringing an important influence to the multiqubit gate performance. A detailed comparison of multi-atom gate performance trapped in a 1D array lattice or 2D triangular lattice would be discussed in Sec. \rm{IV}B.

\section{Multiqubit Toffoli gate Sphere}

\subsection{Toffoli blockade gate sphere}

In a real implementation the three-qubit Toffoli gate has been suggested to an $n$-atom controlled NOT ($C_{n}$NOT) gate extension via common pulses that is independent of $n$ \cite{Isenhower11}. Accompanied by the experimental achievement that mesoscopic Rydberg superatom sphere promising one excitation that shifts all other atoms out of resonance, has been created \cite{Weber15}, we study an extensive multiqubit Toffoli gate characterized by a spherical model, taking the special triangular arrangement of atoms into account, which further proves the scheme validity mainly decided by $d\leq R_b$. Compared to other multiqubit $C_n$NOT schemes based on square lattice \cite{Isenhower11}, cavity QED \cite{Shao07}, quantum circuits \cite{Saeedi13,Wei16} and cavity-waveguide systems \cite{Peng19}, our protocol benefits from an entirely unconstrained spatial arrangement of control atoms inside the Toffoli sphere, enabling a highly efficient implementation of arbitrary-qubit Rydberg Toffoli gate in experiment.

\begin{figure}
\centering
\includegraphics[width=3.4in,height=1.8in]{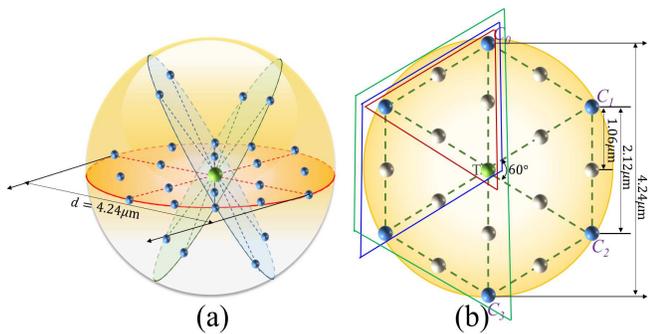}
\caption{(color online). Schematic of the $(n+1)$-qubit Toffoli gate sphere with one central target atom surrounded by maximal $n=46$ control atoms. (a) A global view for the Toffoli sphere in diameter of $d=4.24\mu$m, as same as the blockade radius $R_b$. The minimal two-atom (control-control or control-target) distance is set to be $R_{tc}=R_{cc}=1.06\mu$m, determined by the experimental implementation of the least spacing between two trapped atom lattices \cite{Li08}. The sphere can hold up to 46 control atoms, forming a $(46+1)$-qubit Toffoli gate. (b) A detailed cross-profile description of a regular-hexagon configuration filled with one target (green) and 18 control (blue and yellow) atoms. Noting that a sphere contains only three cross profiles constrained by the minimal two-atom distance $\sim 1.0\mu$m in the experiment. }
\label{sphere}
\end{figure}

As represented in Fig.\ref{sphere}(a) we show an $(n+1)$-qubit Toffoli gate sphere formed by one target atom in the center surrounded by $n$ control atoms, in which the longest spacing between two control atoms is set to be $d=R_b= 4.24\mu$m ensuring a well blockade mechanism at any time between arbitrary atoms inside the sphere. Similar to the preparation of superatom sphere which contains 100-500 atoms or more within a typical diameter of $(3\sim 5)\mu$m \cite{Weber15,Hofmann13} we consider a regular arrangement for atoms inside the Toffoli sphere via 3D lattice technique \cite{Kumar18}, where the nearest spacing between control-control or target-control atoms is $R_{cc}=R_{tc}=1.06\mu$m, depending on the experimental facts that two atoms can be individually trapped in lattices with a minimal spacing of about $1\mu$m \cite{Li08}. 
Therefore, the maximal filling number of control atoms within the Toffoli blockade sphere is $n=46$ in which the behavior of one target atom can be simultaneously manipulated by $46$ surrounding control atoms.

The full distribution of these $46$ control atoms can be classified into three cross profiles, as represented in Fig.\ref{sphere}(b) where the endpoint-control and middle-control atoms are denoted by blue and yellow circles, respectively. Noticing a recent preprint reported the possibility of hundreds of microscopic atomic ensembles for encoded Rydberg qubits via optical tweezer approach also confirming the validity of our protocol \cite{Wang19}. The inclusion of middle-control atoms allows the minimal atomic spacing as small as $\sim1.06\mu$m yet without adding new physics then. Therefore it is reasonable to take some fewer-atom models ($n=2,3,4$) as examples, with endpoint-control atoms only in order to compare gate performances in the presence of different multiqubit atom structures. In the calculation we have ignored the middle-control atoms leading to the nearest spacing of two atoms equal to $R_b/2$.

\subsection{fewer-atom schemes}

A straightforward extension from the $(2+1)$-qubit triangular protocol [see red frame in Fig.\ref{sphere}(b)], is adopting $(3+1)$-quadrangle (blue frame) or $(4+1)$-trapezoid (green frame) configuration, for the purpose of studying the influence of NN interaction or next to next nearest (NNN) interaction. For instance when placing a fourth atom with equal separations to one control atom and the target atom, it leads to the NN spacing $\sqrt{3}$-times bigger than the nearest spacing {\it i.e.} $R_{c-to-c}=\sqrt{3}R_{cc}$ due to its long-range feature. Consequently the NN interaction strength becomes $V_{c-to-c}=\frac{1}{27}V_0$ that may be insufficient for blocking the target excitation. One alternative way to overcome it is reducing the size of gate, allowing the long-range spacing between two NN control atoms $R_{c-to-c}\leq R_b$. To this end the small interaction $V_{c-to-c}$ is still able to realize a blocked excitation threshold for the target atom. Accordingly the same solution can also be applied to the scheme with $(4+1)$ atoms where the NNN distance $R_{c-to-to-c}$ must be smaller than $R_b$.

\begin{widetext}

\begin{figure}
\centering
\includegraphics[width=4.1in,height=5.0in]{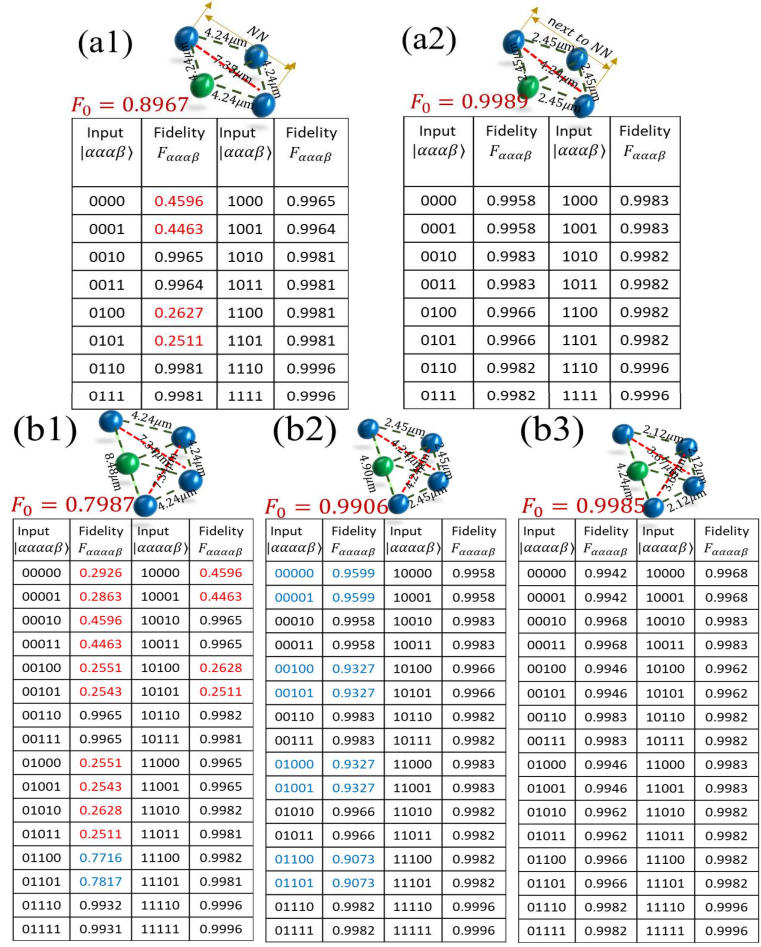}
\caption{(color online). (a1-a2) The truth table of the EF and AF for the gate performance of a $(3+1)$-atom Toffoli scheme where different two-atom distances are used. (a1) $R_{tc}=R_{cc}=R_b=4.24\mu$m and $R_{c-to-c}=\sqrt{3}R_b=7.35\mu$m; (a2) $R_{tc}=R_{cc}=2.45\mu$m and $R_{c-to-c}=R_b=4.24\mu$m. (b1-b3) Same quantities are solved in the case of a $(4+1)$-atom protocol under different interatomic separations: (b1) $R_{tc}=R_{cc}=R_b=4.24\mu$m, $R_{c-to-c}=\sqrt{3}R_b=7.35\mu$m, $R_{c-to-to-c}=2R_b=8.48\mu$m; (b2) $R_{tc}=R_{cc}=2.45\mu$m, $R_{c-to-c}=R_b=4.24\mu$m, $R_{c-to-to-c}=(2/\sqrt{3})R_b=4.90\mu$m; (b3) $R_{tc}=R_{cc}=2.12\mu$m, $R_{c-to-c}=3.67\mu$m, $R_{c-to-to-c}=R_b=4.24\mu$m.}
\label{four}
\end{figure}

\end{widetext}

Relevant calculations depending on a numerical solving of the master equation with same pulses via RK method are presented in Fig. \ref{four} except that the number of input qubit states turns to be $\mathcal{N}=2^4$ or $2^5$ then. For the case of $n=3$ (four atoms), we comparably show the results by setting the nearest spacing $R_{tc}=R_{cc}=R_b$ in (a1) and the NN spacing $R_{c-to-c}=R_b$ in (a2). Clearly, in (a1) the weaker interaction between two NN control atoms ($1/27$ of the blockaded energy), decided by   $R_{c-to-c}=\sqrt{3}R_{b}=7.35\mu$m, is insufficient for preserving the least requirement of strong blockade. Therefore, if atoms $C_0$ and $C_2$ are both prepared in state $|0_c\rangle$ referring to the input states like $|0\alpha0\beta\rangle$, 
the imperfect blockade effect can not suppress the excitation of the target atom,  causing the output fidelity very low, see the table in Fig. \ref{four}(a1) [red texts]. The total AF covering $\mathcal{N}=16$ input states is calculated to be $\mathcal{F}_0=0.8967$. However as expected, if the long-range NN spacing $R_{c-to-c}$ is reduced to be $R_b$ the corresponding nearest neighbor spacing becomes 2.45$\mu$m at the same time, as represented in (a2). With this improvement, we find the AF of gate can return back to a higher level $\sim0.9989$ with all EFs more than 0.99, perfectly verifying the correctness of the blockade threshold as found in section \rm{IIIA}.
Note that this nearest neighbor spacing $\sim2.45\mu$m between two atom lattices is still achievable by current experimental technology.

To further verify the robustness of an $(n+1)$-qubit Toffoli gate sphere, a similar extension is carried out by considering a $(4+1)$-trapezoid structure in which the farthest interatomic distance {\it i.e.} the diameter of sphere is set to be $R_b$ exactly. To this end the addition of extra atoms can safely preserve the blockade condition over the entire Toffoli sphere, never breaking it. From Fig. \ref{four}(b1) to (b3) with decreasing the size of gate by a smaller interatomic spacing, finally our gate attains a maximal AF $\sim 0.9985$ with all EFs above 0.99 when the NNN spacing is lowered to the critical value $R_b$. If not, {\it e.g.} in (b1) only the nearest neighbor distance satisfies the blockade threshold. In that case for $R_{c-to-to-c}=8.48\mu$m, control atoms $C_0$ and $C_3$ are not influenced by blockade effect having a probability to be excited to the Rydberg levels at the same time. 
As a result if state $|0110\beta\rangle$ is prepared (only $C_0$ and $C_3$ can be excited), causing a poor blocking to the excitation of target atom, one still has a relative high fidelity output $0.77\sim0.78$ [blue texts in (b1)]. Otherwise it is remarkable that if other control atoms probably exist on $|0_c\rangle$ simultaneously, an extra breakup of blockade effect leads to a competition between multi-excitation channels among more than two control atoms that would deeply decrease the final EF of the corresponding input states, {\it e.g.} $|0000\beta\rangle$, $|0010\beta\rangle$, $|0100\beta\rangle$ [red texts in (b1)].

Decreasing the length of interatomic distance arises an obvious improvement to the EFs of states $|0\alpha\alpha0\beta\rangle$, as implied in (b2) in which the EFs affected by long-range and multi-excitation obtain an obvious enhancement, reaching above $\sim 0.90$ because the NNN spacing is still $R_{c-to-t-to-c}=4.90\mu$m$>R_b$ in this case. Supported by (b3) that as long as the two farthest endpoint-control atom is separated by $R_b$, all EFs will return back to a high level $>0.99$ ensuring the average gate fidelity $\mathcal{F}_0$ preserving a saturation value $\sim 0.9985$.

Remarkably this value is slightly smaller than the AF values in $(2+1)$- or $(3+1)$-qubit Toffoli gates, primarily determined by the accumulated blockade errors $\mathcal{E}_{b}$ from multiqubit operation, as discussed in section \rm{III}A, because another decay error $\mathcal{E}_{\Gamma}$ is one order of magnitude smaller that can not dominantly affect the fidelity value then. Expectedly for input states $|11\beta\rangle$, $|111\beta\rangle$, $|1111\beta\rangle$ without any blockade error all EFs can be preserved to be a constant $\sim$0.9996 based on our calculations.

\subsection{Comparing with a 1D-array atomic lattice}

\begin{figure}
\centering
\includegraphics[width=3.2in,height=5.0in]{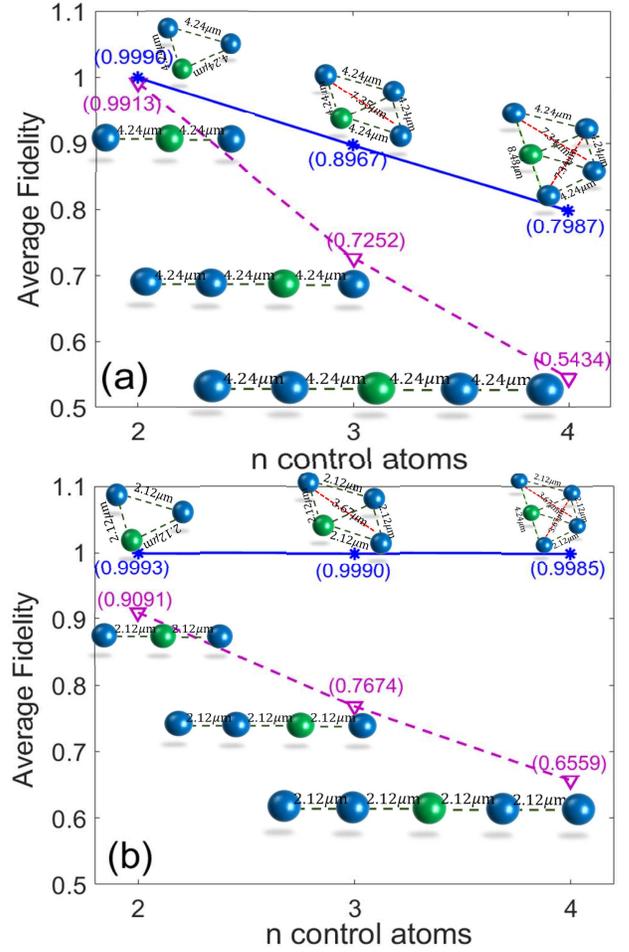}
\caption{(color online). The AF versus the number of control atoms $n=2,3,4$, under a 2D triangular (blue-solid) and an 1D array (red-dashed) protocols. The target and control atoms are denoted by green and blue circles, respectively.  (a) The nearest neighbor spacing is $4.24\mu$m. (b) The nearest neighbor spacing is $2.12\mu$m.}
\label{five}
\end{figure}

In order to show the robustness of the multiqubit Toffoli sphere as compared to an original 1D-array system, Figure \ref{five} presents the final AF of multiqubit gates with the increase of control atom numbers $n$ under two different protocols, where the nearest spacing is set to be (a) $R_b$ and (b) $R_b/2$, respectively. In (a) by utilizing a 1D-array protocol, due to the longer-range imperfect blockade mechanism directly increasing with $n$ the AF suffers from a straightforward fall, attaining as low as 0.5434 if $n=4$. Nevertheless, our novel structure can achieve a higher fidelity for $n=4$ owing to its well preserved blockade effect. Because in the 1D model the strong blockade effect between two nearest control atoms on one side may cause a competition of excitation in themselves, difficult for blocking the target excitation. If it happens the target atom will naturally change its final state by experiencing an excitation and de-excitation process, breaking the input-output procedure.

Fortunately, while reducing the nearest neighbor spacing into $R_b/2$ our scheme manifests as a perfect multiqubit Toffoli gate that can strongly preserve the AF at a very high level no matter what the number of control atoms is. It mainly benefits from the Toffoli sphere model as similar as a superatom blockaded sphere, in which two arbitrary atoms suffer from a blockaded energy once its diameter is kept to be smaller than the blockade radius $R_b$. However this solution can not work for the case of a 1D array lattice as confirmed by (b) in which the AF still acquires a big decrease with $n$, because the imperfect blockade remains between longer-range control-control atoms based on a 1D array lattice.

\section{Conclusion}

The development of single superatom preparation in a three-dimensional lattice system may provide a clean platform for predominant gate operation in quantum computation. We propose a spherical {\it (n+1)}-qubit Toffoli gate protocol by manipulating the behavior of one central target atom with $n$ surrounding control atoms in the sphere, essentially enabled by the strong Rydberg blockade. This mechanism was widely utilized in other quantum logic gates with Rydberg atoms \cite{Moller08,Theis16,Zeng17}. However, differing from the previous works here a simpler and scalable spherical configuration is considered ensuring a perfect preservation for any two-atom strong blockade, once the diameter of sphere is equal to or smaller than the two-atom blockade radius. Therefore, the only constraint determining the maximal filling atomic number inside the sphere, is the least spacing between two trapped-atom lattices, characterized by the nearest neighbor interaction. So far the period of lattice depending on the angle of two forming beams can be easily adjusted to a desired value in experimental setup \cite{Fallani05}, from $0.96\mu$m$\sim11.2\mu$m in one second \cite{Li08}. Hence by considering the nearest spacing as small as 1.06$\mu$m, the maximal filling number contains one target atom plus 46 surrounding control atoms, forming a {\it (46+1)}-qubit Toffoli gate sphere. The trapping technology for developing large-scale neutral atom qubit systems has recently reported via optical tweezers in ytterbium Rydberg atoms, promising for a 3D multi-array Rydberg atom confinement \cite{Wilson19}.

We comparably demonstrate the high-quality performance of {\it (2+1)}-, {\it (3+1)}-, and {\it (4+1)}-qubit Toffoli gates by exactly solving the output probabilities via five common $\pi$ pulses under the variation of relative blockade strengths, which confirms the importance of Rydberg blockade threshold for realizing a higher average gate fidelity. Detailed analyses for the quality of a {\it (2+1)}-qubit gate, taking into account realistic experimental parameters, show that the gate errors are mainly attributed to the imperfect blockade strength, the spontaneous decay from Rydberg levels as well as the imperfect ground state preparation, which have a quantitatively agreement with the previous theoretical predictions. In contrast to the typical 1D-array schemes, our Toffoli sphere protocol enables a well preservation for a high gate fidelity with the increase of $n$, ensuing a flexible extension to realize an arbitrary multiqubit fast gate operation as well as the production of Rydberg-mediated multi-particle entanglement of tens of atoms in a higher-dimensional atom array.

\bigskip

\acknowledgements

This work was supported by the National Natural Science Foundation of China under Grants Nos. 11474094, 91950112, 11174081,11104076,  the Science and Technology Commission of Shanghai Municipality under Grant No. 18ZR1412800, the National Key Research and Development Program of China under Grant No. 2016YFA0302001, and the Academic Competence Funds for the outstanding doctoral students under Grant No. YBNLTS2019-023.

\appendix

\end{document}